\documentclass[prb,showpacs,preprintnumbers,twocolumn,amsmath,amssymb]{revtex4}
\usepackage{graphicx}%
\usepackage{color}
\usepackage{bm}
\usepackage{amsmath}
\usepackage{braket}
\renewcommand\Im{\operatorname{Im}}
\renewcommand\Re{\operatorname{Re}}
\begin{document}

\title{\textit{Ab-initio} analysis of plasmon dispersion in sodium under pressure}

\author{Julen Iba\~{n}ez-Azpiroz$^{1,2}$, Bruno Rousseau$^4$, Asier Eiguren$^{1,2}$, Aitor Bergara$^{1,2,3}$}
\address{$^{1}$Materia Kondentsatuaren Fisika Saila, Zientzia eta
Teknologia Fakultatea, Euskal Herriko Unibertsitatea, 644
Postakutxatila, 48080 Bilbao, Basque Country, Spain}
\address{$^{2}$Donostia International Physics Center (DIPC), Paseo Manuel de Lardizabal 4, 20018 Donostia/San Sebastian, Spain}
\address{$^3$Centro de F\'{i}sica de Materiales CFM - Materials Physics Center MPC, Centro Mixto CSIC-UPV/EHU,
Edificio Korta,
Avenida de Tolosa 72, 20018 Donostia, Basque Country, Spain}
\address{$^4$D\'epartement de physique et Regroupement qu\'eb\'ecois sur les mat\'eriaux de
pointe (RQMP),Universit\'e de Montr\'eal,
C. P. 6128 Succursale Centre-ville,
Montr\'eal (Qu\'ebec),
H3C 3J7 Canada}

\date{\today}


\begin{abstract}
We present an \textit{ab-initio} study of 
the electronic response function of sodium in 
its 5 known metallic phases from 0 to 180 GPa 
at room temperature. 
The considered formalism is based on a
interpolation scheme within time-dependent density functional theory 
that uses maximally localized Wannier functions,
providing an accurate sampling of the reciprocal space. 
Besides showing an excellent agreement with 
inelastic X-ray scattering experiments~\cite{loa_plasmons_2011,Mao20122011},
our calculations reveal that the drastic decrease of the optical reflectivity
recently measured in the high pressure phases oP8 and tI19~\cite{anomalous-pnas} 
is associated to
a new low-energy plasmon
arising from collective interband excitations.
Additionally, our calculations predict the existence of an anisotropic interband 
plasmon in the stability pressure range of fcc Na (65 to 105 GPa).

\end{abstract}
\maketitle

\section{INTRODUCTION}

Sodium is one of the closest realizations of the 
free-electron gas that can be found in solid-state materials.
Under compression, however, the simple metal behavior of
Na is deeply modified by non-free-electron-like 
features of its band structure; these are
mainly associated to the increasing
electronic hybridization 
of the chemical bondings  
and the strong
non-local character of the pseudopotential, 
among other causes.~\cite{as-ne-sodium}
As a consequence, sodium
under pressure 
develops
a variety of
unexpected processes
including phase transitions
to extremely complex structures~\cite{structural-science},
loss of the metallic character~\cite{ma_transparent_2009}
or anomalies in the optical response
to external fields~\cite{anomalous-pnas,na-calc}.
These remarkable phenomena  
challenge the
classical viewpoint
that pressure should make simple metals
even simpler. 
 
According to room temperature X-ray diffraction 
experiments,~\cite{poster,structural-science,ma_transparent_2009,na-calc} 
sodium undergoes a series of structural
phase transformations from 0 to $\sim$180 GPa
before it experiences a metal-insulator
transition that suppresses its metallic properties. 
Over this wide pressure range, 
sodium first adopts the bcc
structure (0 to 65 GPa), 
followed by the fcc
(65 to 105 GPa), 
the cI16 (105 to 118 GPa),
the oP8 (118 to 125 GPa)
and the tI19 (125 to $\sim$180 GPa) configurations.~\cite{exotic-li-so}
Similar to what happens in other \textit{a priori} simple metals,
such as lithium or calcium,
the metallic properties of sodium at high pressures 
are strongly modified. 
In particular, the high and uniform reflectivity
(characteristic of good metals) 
of the bcc, fcc and cI16 phases
has been measured to drop drastically 
in the high pressure phases oP8 and tI19,
accompanied by a decrease of the metallic character.~\cite{anomalous-pnas}
The origin of such behavior may lie
on the emergence of low-energy interband plasmons 
arising from the 
increasing localization of the valence electrons,~\cite{as-ne-lithium}
as it is the case of Li and Ca.~\cite{silkin_strong_2007,calcium-ion}

In this paper, we perform a detailed \textit{ab-initio} analysis
of the electronic properties of Na in a  
pressure domain ranging from 0 to 180 GPa, covering
all the metallic phases of sodium at room temperature.
In the spirit of Ref. \onlinecite{magnon},
we employ a formalism based on Wannier 
interpolation to obtain the electronic linear response
within time-dependent 
density functional theory (TDDFT).  
This approach allows us to 
perform a very accurate sampling of
the relevant functions in
reciprocal space, which
is essential for 
describing the low-energy   
excitations that 
determine the response function. 
As it will be shown,
our calculations predict the existence of a very low-energy
plasmon in the oP8 and tI19 phases of sodium that
explains the anomalous 
optical properties measured
at the corresponding pressures.~\cite{anomalous-pnas}
In addition, our analysis of the lower pressure
phase (fcc) indicates that sodium develops an anisotropic interband plasmon 
originating from a band structure effect along
the $\Gamma$L direction.

The paper is organized as follows. In Sec.  
\ref{sec:na-theory}, the Wannier-based interpolation
scheme for calculating the plasmon dispersion 
within TDDFT is presented, 
along with the details regarding 
the ground state \textit{ab-initio} calculations. 
The method is applied to calculate
the plasmon dispersion in the metallic phases
of Na; 
results and their analysis are presented in Sec. \ref{sec:results}.
A summary and conclusions are presented in Sec. \ref{sec:conclusions}.
Unless otherwise stated,
atomic units are used throughout the work 
($\hbar=m_{e}=e^{2}=4\pi\epsilon_{0}=1$).

\section{THEORETICAL FRAMEWORK}
\label{sec:na-theory}

In this section we review the 
formalism for calculating the dispersion of collective charge excitations 
in solids within TDDFT.~\cite{tddft-1,tddft-2}
The key ingredient for such a task is the interacting response function,
which describes 
the variation of the 
electronic density induced by an external time-dependent potential. 
The expression for this quantity is given by
\begin{multline}\label{pl-response-q}
\chi_{\textbf{q}}(\textbf{r},\textbf{r}^{\prime},\omega)=
\chi_{\textbf{q}}^\text{KS}(\textbf{r},\textbf{r}^{\prime},\omega)+
\frac{1}{\Omega}\int d\textbf{r}_{1}\frac{1}{\Omega}\int d\textbf{r}_{2}\\
\times\chi_{\textbf{q}}^\text{KS}(\textbf{r},\textbf{r}_{1},\omega)
K_{\textbf{q}}(\textbf{r}_{1},\textbf{r}_{2},\omega)
\chi_{\textbf{q}}(\textbf{r}_{2},\textbf{r}^{\prime},\omega),
\end{multline}
where $\Omega$
denotes the unit cell volume. 
In Eq. \ref{pl-response-q}, 
$\chi_{\textbf{q}}^\text{KS}(\textbf{r},\textbf{r}^{\prime},\omega)$ represents
the non-interacting Kohn-Sham (KS) response function for a given frequency $\omega$ and 
momentum \textbf{q}.
The term 
$K_{\textbf{q}}(\textbf{r},\textbf{r}^{\prime},\omega)$ 
is the kernel 
that takes into account the electron-electron interactions:
\begin{equation}\label{eq:kernel}
K_{\textbf{q}}(\textbf{r},\textbf{r}^{\prime},\omega) = \frac{e^{2}}{|\textbf{r}-\textbf{r}^{\prime}|} + f_{\textbf{q}}^\text{xc}(\textbf{r},\textbf{r}^{\prime},\omega).
\end{equation}
The first term in Eq. \ref{eq:kernel} is the Coulomb interaction 
associated to the electronic charge, while  
$f_{\textbf{q}}^\text{xc}(\textbf{r},\textbf{r}^{\prime},\omega)$ contains 
the exchange and correlation effects.
In the present work, 
$f_{\textbf{q}}^\text{xc}(\textbf{r},\textbf{r}^{\prime},\omega)$ 
has been approximated within 
the PZ-LDA parametrization.~\cite{perdew-zunger,ldalinear}

In terms of the single-particle KS orbitals $\psi_{n {\bf k}}({\bf r})$, 
the expression of the non-interacting KS response function 
is given by~\cite{adler-response}
\begin{multline}\label{ks-rf}
\chi_{\textbf{q}}^\text{KS}({\bf r,r'},\omega) 
	=
	\sum_{n_1,n_2} \sum_{\bf k}^\text{1BZ}
	\frac{f(\xi_{n_1{\bf k}})-f(\xi_{n_2{\bf k+q}})}
	{ \omega+\xi_{n_1{\bf k}} -\xi_{n_2{\bf k+q}}+i\delta}\\
	\times\psi_{n_1 {\bf k}}^{*}({\bf r})
	\psi_{n_2 {\bf k+q}}({\bf r})
	\psi_{n_2 {\bf k+q}}^{*}({\bf r'})
	\psi_{n_1 {\bf k}}({\bf r'}),
\end{multline}
where $\delta$ is an infinitesimal positive 
parameter ensuring causality. 
In Eq. \ref{ks-rf}, \textbf{k} is constrained to the first Brillouin zone (1BZ),
$n_1$ and $n_2$ are band indices, $\xi_{n{\bf k}}=\epsilon_{n{\bf k}} - \mu$   
with $\epsilon_{n{\bf k}}$ a KS eigenvalue and $\mu$ the chemical potential,
and $f(\xi)$ represents the Fermi-Dirac distribution function.

\subsection{Maximally localized Wannier functions}

An appropriate basis set for calculating  
$\chi_{\textbf{q}}^\text{KS}({\bf r,r'},\omega)$
can be constructed in terms of maximally localized Wannier functions (MLWFs).~\cite{magnon} 
The relationship between the Wannier states and the KS states is
given by
\begin{align}
\label{eq:wann}
W_{n}({\bf r-R}) &= \frac{1}{\sqrt{N}}\sum_{m\bf k} e^{-i\bf k\cdot R}
		\psi_{m\bf k}({\bf r}) \textbf{U}_{mn}({\bf k}).
\end{align} 
In Eq. \ref{eq:wann}, \textbf{R} denotes a periodic 
lattice vector while $\textbf{U}({\bf k})$ is a 
unitary matrix in band indices.~\cite{marzari1997}
In practical calculations, 
the sum in Eq. \ref{eq:wann} is over a 
\textbf{k}-point mesh 
that must accurately reproduce the electronic band structure
of the system.
The first principles calculation of the ground-state orbitals
in that mesh usually requires a substantial computational effort. 
However, once the Wannier functions have been constructed via Eq. \ref{eq:wann},
quantities such as eigenvalues, 
eigenfunctions or the above mentioned unitary matrices can
be interpolated into a much finer \textbf{k}-point mesh 
using a computationally inexpensive fast Fourier transform algorithm.

The non-interacting response function defined in Eq. \ref{ks-rf}
can be cast into the following form
with the aid of the Wannier functions,
\begin{multline}
\label{eq:chi-q}
\chi_{\textbf{q}}^{\text{KS}}({\bf r,r'},\omega)  
	= \sum_{IJ}
	\Big[B_{I,\textbf{q}}({\bf r})\Big]~
	\bm{\chi}^{\text{KS}}_{IJ}({\bf q},\omega) ~
	\Big[B_{J,\textbf{q}}({\bf r})\Big]^*,\\
	 I \equiv \{m_1,m_2,{\bf R}\},\quad J \equiv \{m_3,m_4,\bf R'\},
\end{multline}
where the lattice periodic functions $\{B_{I,\textbf{q}}({\bf r})\}$,
which we will henceforth refer to as the bare basis, 
are given by 
\begin{multline}
\label{eq:bare-basis}		
		B_{I,\textbf{q}}({\bf r}) =
		\frac{\Omega}{N}\sum_{\bf R^{\prime}}
		e^{i\bf q\cdot (R^{\prime}-r)}\\
		\times W^{*}_{m_1}({\bf r-[R^{\prime}-R]}) 
		W_{m_2}({\bf r-R^{\prime}}).
\end{multline}
We notice that the bare basis of Eq. \ref{eq:bare-basis}
has a trivial dependence on the external momentum \textbf{q},
as it only enters in the exponential factor. 
Thus, once the product 
$W^{*}_{m_1}({\bf r-[R^{\prime}-R]}) 
		W_{m_2}({\bf r-R^{\prime}})$
is calculated and stored, the bare basis for
different momenta is straightforwardly obtained.
This allows to efficiently map the evolution of
the response function as a function of \textbf{q},
which is of critical importance when analyzing 
the dispersion of plasmons.

The coefficients $\bm{\chi}^{\text{KS}}_{IJ}({\bf q},\omega)$ entering Eq. \ref{eq:chi-q}
do not explicitly depend on
the Wannier functions, but only on the unitary matrices,
\begin{multline}
\label{eq:response-spin-indp}
	\bm{\chi}^{\text{KS}}_{IJ}({\bf q},\omega) =
	\frac{1}{\Omega} \sum^\text{1BZ}_{\bf k} 
	e^{i\bf k\cdot (R-R')}\\
	\times\sum_{n_1,n_2} 
	\frac{f(\xi_{n_1{\bf k}})-f(\xi_{n_2{\bf k+q}})}
	{ \hbar\omega+\xi_{n_1{\bf k}} -\xi_{n_2{\bf k+q}}+i\delta}\\
	\times\textbf{U}_{n_1m_1}({\bf k})
	\textbf{U}^\dagger_{m_2n_2}({\bf k+q})
	\textbf{U}^\dagger_{m_3n_1}({\bf k})
	\textbf{U}_{n_2m_4}({\bf k+q}).
\end{multline}
In practice, we have collected all the \textbf{k}-dependent
quantities into the above coefficients, given that the bare basis functions are 
\textbf{k}-independent.
It is noteworthy that all the ingredients in Eq. \ref{eq:response-spin-indp}
can be calculated on a fine \textbf{k} mesh using 
the Wannier interpolation scheme, allowing a very fine sampling
of $\bm{\chi}^{\text{KS}}_{IJ}({\bf q},\omega)$.

We notice that in contrast to the bare basis, the dependence
of $\bm{\chi}^{\text{KS}}_{IJ}({\bf q},\omega)$ on 
the external momentum is not trivial as it involves terms
like $\textbf{U}_{mn}({\bf k+q})$,  $\epsilon_{n{\bf k+q}}$ and $f_{n{\bf k+q}}$. 
In principle, these terms should be calculated \textit{ab-initio}
for each different \textbf{q}.
However, if the point ${\bf k+q}$ lies inside the 
interpolated grid, then the \textbf{q}-dependent quantities of 
Eq. \ref{eq:response-spin-indp} are directly available.
In practice, Wannier interpolation allows to
consider such fine \textbf{k} meshes so that choosing \textbf{q}
inside the interpolated grid does not represent a limitation.

From the computational point of view, it is of practical interest to express 
the central equation describing the interacting 
response function (Eq. \ref{pl-response-q}) 
as a matrix equation.
Let us consider the kets 
$\ket{B_{I,\textbf{q}}}$ associated to the bare basis functions 
and represent the KS response function as
\begin{align}
\label{eq:ks-operator}
\hat{\boldsymbol{\chi}}_{\textbf{q}}^\text{KS}=
\sum_{IJ}\ket{B_{I,\textbf{q}}}\bm{\chi}^{\text{KS}}_{IJ}({\bf q},\omega)\bra{B_{J,\textbf{q}}}.
\end{align}
In this way, Eq. \ref{pl-response-q} can be written as a matrix equation,
\begin{equation}
\label{eq:self-cons}
\hat{\boldsymbol{\chi}}_{\textbf{q}} =   \left(\hat{\boldsymbol{1}} -  
	\hat{\boldsymbol{\chi}}_{\textbf{q}}^\text{KS} \cdot \hat{\boldsymbol{K}}_{\textbf{q}}\right)^{-1}\cdot	\hat{\boldsymbol{\chi}}_{\textbf{q}}^\text{KS}  .
\end{equation}
The above matrix equation 
must be truncated into a finite size problem in order to 
be solved numerically. 
Given that the basis functions 
$B_{I,\textbf{q}}({\bf r})$ are not linearly independent,
it is ineffective to compute all the coefficients 
$\bm{\chi}^{\text{KS}}_{IJ}({\bf q},\omega)$
independently.
Instead, it is convenient to establish a minimal basis set 
that describes the essential physics of the problem.

\subsubsection{Crystal local field effects}

Plasmons are described as peaks in $\Im\hat{\bm{\chi}}_{{\bf q}}(\omega)_{\textbf{00}}$.~\cite{plasmon-ref}
Therefore, it is sensible to include the function 
$\ket{1}$ (i.e., a plane wave with
$\textbf{G}=\textbf{0}$) in the minimal basis set. 
In addition, crystal local field effects (CLFE)
often play an important role in
determining the plasmon dispersion;
usually, 
wave vectors other than $\textbf{q}$
are 
needed for describing
the spatial variation of external fields
inside the solid due to the inhomogeneity of the system.~\cite{sturm,adler-response}
Generally, the use of a finite number of $\textbf{G}$ vectors 
properly describes the CLFE. 
Therefore, we include plane waves 
\begin{equation}
\label{eq:plane-wave}
\frac{1}{\sqrt{\Omega}}e^{i\textbf{G}\cdot\textbf{r}}\rightarrow\ket{\textbf{G}}
\end{equation}
in the minimal basis set; the number of 
$\textbf{G}$ vectors to include is a parameter
to be converged. 
Thus, the minimal basis should span the same functional space as the set of functions:
\begin{equation}
\label{eq:basis-space}
\ket{g_{i}} \in \left\{ \ket{1}, \left\{\ket{\textbf{G}}\right\}  \right\}.
\end{equation}
We define the minimal basis functions as
\begin{equation}
\label{eq:optimal-basis}
\ket{b_{i}} = \sum_{j} \ket{g_{j}} g_{ij},\;\; \braket{g_{j}|b_{i}}=g_{ij}, \; \;
\braket{b_{i}|b_{j}}=\delta_{ij}.
\end{equation}
In this way, the self-consistent Eq. \ref{eq:self-cons} 
regarding the interacting response function
can be solved 
by projecting the relevant functions into the minimal basis set,
\begin{equation}
\begin{split}
\left[ \hat{\boldsymbol{\chi}}_{\textbf{q}}  \right]_{ij} & 
\simeq \sum_{l} 
\left[ \left(\hat{\boldsymbol{1}} -  
	\hat{\boldsymbol{\chi}}_{\textbf{q}}^\text{KS} \cdot \hat{\boldsymbol{K}}_{\textbf{q}}\right)^{-1} \right]_{il} \left[ \hat{\boldsymbol{\chi}}_{\textbf{q}}^\text{KS}  \right]_{lj},
\end{split}
\end{equation}
where the latin indices refer to the
functions $b_{i}({\bf r})$.
Finally, for future analysis
of plasmon-related properties,
it is convenient to write down the expression
of the inverse dielectric matrix in this subspace,
\begin{equation}
\left[ \hat{\boldsymbol{\epsilon}}^{-1}_{\textbf{q}}  \right]_{ij}  \simeq 
\delta_{ij}+\sum_{l}\left[\hat{\boldsymbol{K}}_{\textbf{q}}\right]_{il}\cdot
\left[\hat{\boldsymbol{\chi}}_{\textbf{q}}\right]_{lj}
.
\end{equation}
   
\subsection{Computational details}

The DFT calculations 
for the ground state eigenvalues and eigenfunctions
have been performed using the QUANTUM-ESPRESSO package~\cite{espresso}, 
with plane waves as the basis set for the expansion of the 
KS orbitals.
The cutoff energy used 
to determine the size of the plane wave basis has been 120 Ry.
The exchange-correlation energy has been approximated within 
the LDA parametrization~\cite{ldalinear,perdew-zunger}
and the 1BZ
has been sampled on a $12\times12\times12$
\textbf{k}-point mesh~\cite{MParck}.

The electron-ion interaction has been modeled  
considering a non-relativistic pseudopotential for Na 
generated with the OPIUM code~\cite{opium}
and tested with all-electron calculations 
performed with the ELK code~\cite{elk}.
We have included $2s^{2}2p^{6}3s^{1}$ states in the valence in order
to properly describe short range effects induced by pressure.

The postprocessing step for obtaining the 
MLWFs has been done using the
WANNIER90 code~\cite{wannier90}.
We have taken into account all bands up to 35 eV above the Fermi level. 
Once the MLWFs have been constructed, the necessary ingredients
for calculating the interacting response function, namely eigenvalues, occupation
factors and rotation unitary matrices, have been
interpolated on a fine $80\times80\times80$ \textbf{k}-point mesh.
Regarding CLFE, the use of 3 reciprocal lattice shells
has yielded converged results in all the phases. 
In our calculations, we find that CLFE have a 
minor effect on the plasmon energy, 
which is modified by less than $2\%$
by the inclusion of CLFE in all cases.

The effects of compression have been simulated by
reducing the lattice parameter. For the bcc and fcc
configurations, we have used the experimental parameters
extracted from the equation of state of sodium at 
the corresponding pressures~\cite{equation-of-state}. 
For the cI16 and oP8 phases,
we have considered the lattice parameters reported in high pressure 
experiments~\cite{anomalous-pnas,structural-science}.
For the tI19 phase, sodium adopts an incommensurate 
host-guest configuration 
with 16 host atoms distributed in a tetragonal 
bcc structure.~\cite{structural-science}
We have modeled this incommensurate phase
by the closely related commensurate tI20 structure, 
containing 20 atoms per unit cell~\cite{anomalous-pnas}.
Due to the wide stability pressure range of tI19 (125 to 180 GPa),
we have analyzed the evolution of its electronic properties at
different pressures. As for this structure there is no accessible experimental 
lattice parameters at the present time, we have used the theoretically calculated ones:
$a=$ 6.59, 6.46, 6.34 a.u. and $c=$ 3.65, 3.54, 3.42 a.u. 
for 125, 150 and 180 GPa, respectively.

Regarding the calculation
of the KS response function, we have 
explicitly computed the absorptive part:
\begin{multline}\label{im-ks-rf}
\Im\chi_{\textbf{q}}^\text{KS}({\bf r,r'},\omega) 
	=
	\sum_{n_1,n_2} \sum_{\bf k}^\text{1BZ}
	f(\xi_{n_1{\bf k}})-f(\xi_{n_2{\bf k+q}})
		\\
	\times\psi_{n_1 {\bf k}}^{*}({\bf r})
	\psi_{n_2 {\bf k+q}}({\bf r})
	\psi_{n_2 {\bf k+q}}^{*}({\bf r'})
	\psi_{n_1 {\bf k}}({\bf r'})\\
	\times
	\delta \left( \hbar\omega+\xi_{n_1{\bf k}} -\xi_{n_2{\bf k+q}} \right),
\end{multline}
while the reactive part $\Re\chi_{\textbf{q}}^\text{KS}({\bf r,r'},\omega)$
has been obtained applying the Kramers-Kronig relations~\cite{mahan}.
The Dirac delta distribution appearing in Eq. \ref{im-ks-rf} 
has been approximated by a gaussian function.
At each \textbf{k}-point, the width of 
the gaussian function has been 
adapted to the
steepness of the integrand of Eq. \ref{im-ks-rf}.~\cite{magnon}
The necessary band gradients have been straightforwardly obtained 
using Wannier interpolation~\cite{k-gradients}.

\section{RESULTS AND DISCUSSION}
\label{sec:results}

In this section we present the calculated electronic properties 
of sodium using the formalism introduced in Sec. \ref{sec:na-theory}.
Our calculations cover a pressure range from 0 to 180 GPa.  
We divide the analysis in two different parts, namely the 0-105 GPa range,
analyzed in Sec. \ref{sec:na-bcc-fcc}, and the
105-180 GPa range, analyzed in Sec. \ref{sec:na-complex-phases}. 
In the first one, sodium adopts
the bcc and fcc structures, which are considerably simpler 
than the ones arising above 105 GPa.

\subsection{Simple phases of sodium}
\label{sec:na-bcc-fcc}

\subsubsection{The bcc phase (0-65 GPa)}
\label{sec:bcc}

\begin{figure}[t]
\centering
\includegraphics[width=0.49\textwidth]{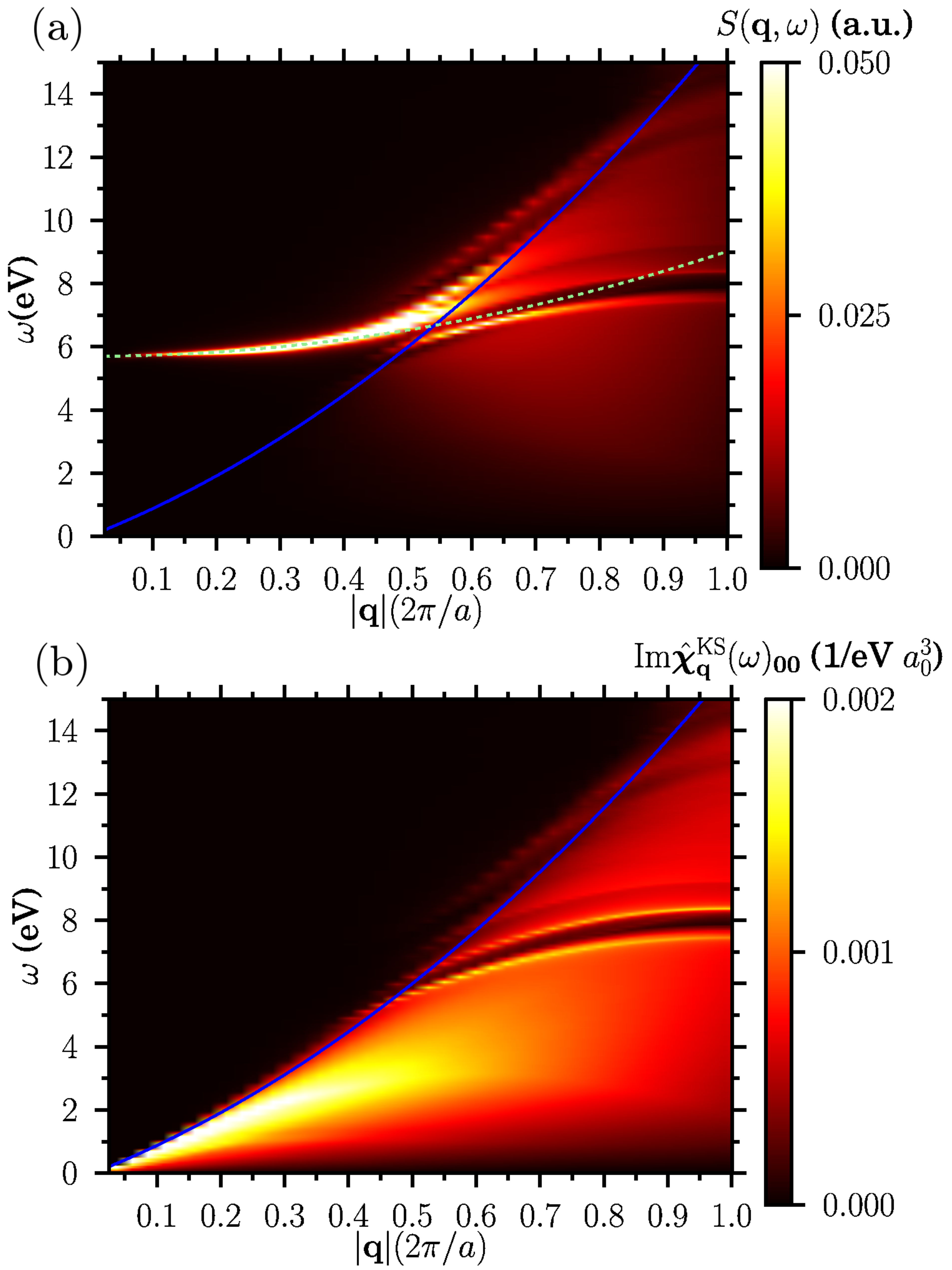}
\caption{(color online)  (a) and (b) show the dynamical structure factor and 
imaginary part of the KS response function
of bcc Na at ambient pressure
along the $\Gamma$P direction, respectively. 
In (a), the dashed (green) line depicts the plasmon dispersion
in the RPA free-electron model (Eq. \ref{eq:pl-disp-RPA}),
while the solid (blue) line in (a) and (b) represents the boundary 
$ q^{2}/2m_{\text{eff}} + qv_{F}$
of the intraband electron-hole excitations.}
\label{fig:bcc-dyn}
\end{figure}

We begin analyzing the electron-hole and collective excitations of
bcc Na at ambient pressure. 
In Fig \ref{fig:bcc-dyn}a, we show the calculated 
dynamical structure factor, 
\begin{equation}
\label{eq:dyn-str}
S(\textbf{q},\omega)=-\frac{|\textbf{q}|^{2}}{4\pi^{2}}\Im 
\hat{\bm{\epsilon}}_{\textbf{q}}^{-1}(\omega)_{\textbf{00}},
\end{equation} 
while
the electron-hole excitation spectrum is
analyzed in Fig. \ref{fig:bcc-dyn}b, where we show the calculated 
$\Im \hat{\bm{\chi}}^\text{KS}_{\textbf{q}}(\omega)_{\textbf{00}}$.

Fig \ref{fig:bcc-dyn} shows a quantitative agreement between our calculations 
and the predictions of the free-electron model.
For low values of the momentum, the calculated plasmon dispersion (Fig. \ref{fig:bcc-dyn}a)
follows the classical RPA expression~\cite{sturm},
\begin{equation}
\label{eq:pl-disp-RPA}
E_{p}(q)=\omega_{p}+\frac{\alpha_{RPA} }{m_{\text{eff}}}q^{2},
\end{equation}
with $m_{\text{eff}}$ the effective electron mass,
$\alpha_{RPA}=\frac{3}{5}\frac{E_{F}}{m_{\text{eff}}\omega_{p}}$ a dimensionless
dispersion constant, $E_{F}$ 
the Fermi energy and
$\omega_{p}=\sqrt{4\pi n}/m_{\text{eff}}$ the 
plasmon energy in the 
free-electron model, where
\textit{n} is the valence electron density.
At ambient pressure, the free-electron-like plasmon
for $\textbf{q}\rightarrow0$ is expected to be located around 
$E_{p}(q\rightarrow0)=\omega_{p}\simeq5.8$ eV, in 
good agreement with
the calculated value $\sim 5.7$ eV.
For finite values of the momentum, 
\begin{figure*}[t]
\centering
\includegraphics[width=0.7\textwidth]{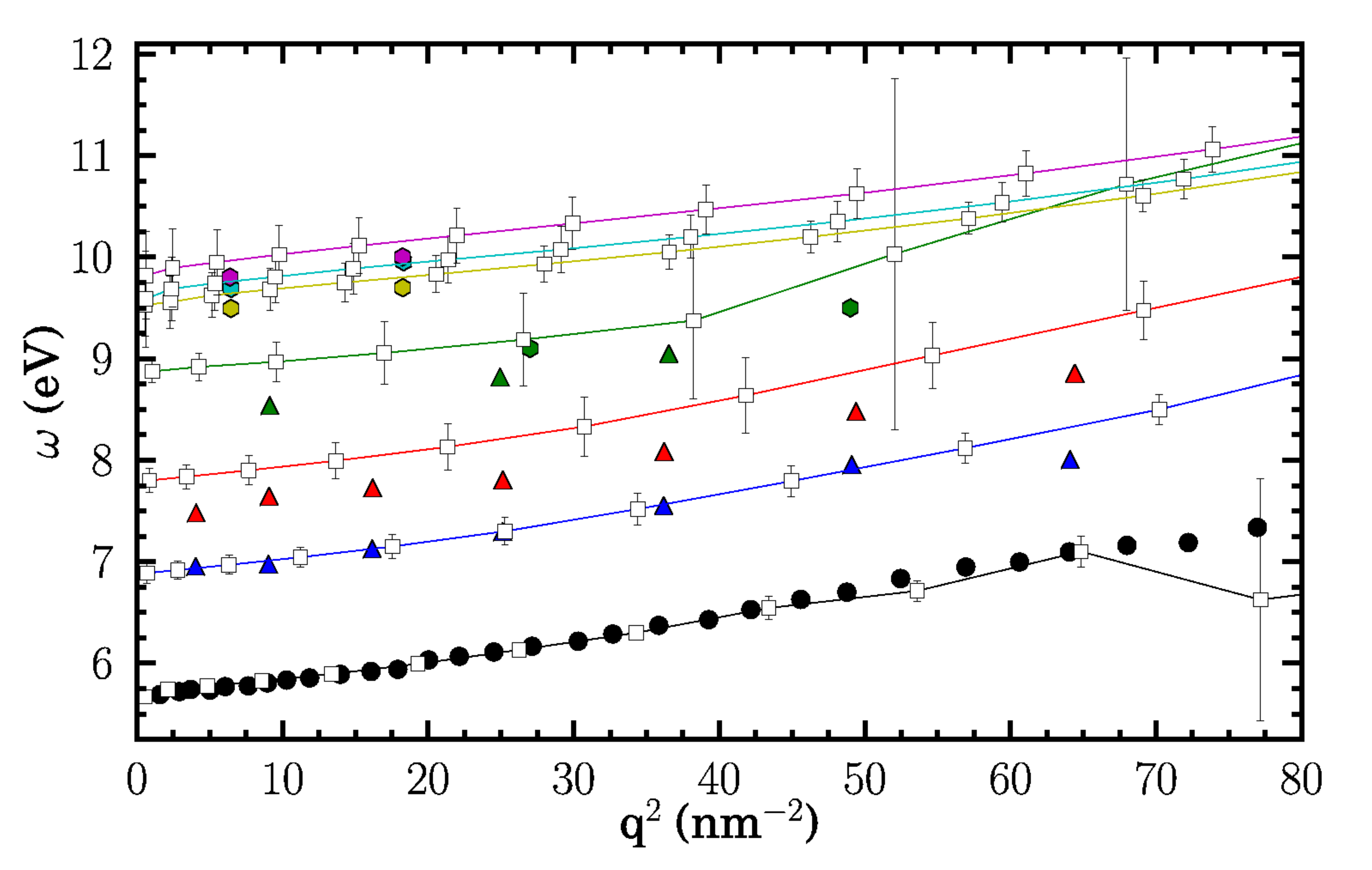}
\caption{(color online) Plasmon dispersion $E_{p}(q^{2})$ of bcc and fcc Na
for 0 (black), 8 (blue), 16 (red), 43 (green), 
75 (yellow), 87 (cyan) and 97 (purple) GPa. Empty squares represent 
our \textit{ab-initio} results, with
the calculated plasmon linewidth
indicated by the markers.
The full lines are simple guides to the eye.
Circles denote
experimental EELS data taken from Ref. \onlinecite{bcc-0gpa},
while triangles and hexagons denote experimental
IXS data taken from Refs. \onlinecite{loa_plasmons_2011} and \onlinecite{Mao20122011},
respectively.
The experimental energy resolution of
the circles, triangles and hexagons are 
0.16 eV, 0.6 eV and 0.1 eV, respectively.
}
\label{fig:pl-disp-bcc}
\end{figure*}
Fig. \ref{fig:bcc-dyn}a displays a smooth parabolic dispersion
of the plasmon until it decays
into the electron-hole continuum at around $6.3$ eV.
This is clear from Fig. \ref{fig:bcc-dyn}b, where 
the border of the electron-hole continuum can be inferred
from the free-electron model prediction,
\begin{equation}\label{eq:rpa-branch}
\omega\leq \dfrac{q^{2}}{2m_{\text{eff}}} + qv_{F},
\end{equation}
with $v_{F}=\sqrt{2E_{F}/m_{\text{eff}}}$ the Fermi velocity.
The above dispersion is shown as a solid blue line
in Fig. \ref{fig:bcc-dyn}b, matching 
very well the calculated border.

In Fig. \ref{fig:pl-disp-bcc} we present the calculated plasmon 
dispersion in the bcc phase at 0, 8, 16 and 43 GPa 
extracted from the position of the peaks in the energy-loss function 
at these pressures. 
We have plotted the plasmon energies as a function of 
$q^{2}$, since we expect the parabolic dependence of 
Eq. \ref{eq:pl-disp-RPA}. 
Overall, we find that the dispersion 
is indeed very close to parabolic at all pressures, 
though the results at 0 and 43 GPa show a slight 
slope change at $q^{2}\sim 40$ nm$^{-2}$ and $q^{2}\sim 65$ nm$^{-2}$, 
respectively.
As revealed by the calculated linewidth,
which increases up to $\sim$2 eV at the mentioned momenta,
the change of slope 
is due to the damping of the plasmon,
which ceases to be a well defined collective excitation
at those points.

For comparison, in Fig. \ref{fig:pl-disp-bcc} we have included experimental data
obtained  by electron energy-loss spectroscopy (EELS) at ambient pressure~\cite{bcc-0gpa} 
and inelastic X-ray scattering (IXS) at higher pressures~\cite{loa_plasmons_2011,Mao20122011}.
As it can be appreciated, 
our results are essentially in agreement with the 
experimental data.
In the case of 0 and 8 GPa, both the calculated energies and dispersion slopes 
are practically identical to the IXS data of Ref.~\onlinecite{loa_plasmons_2011}. 
At 16 and 43 GPa, the calculated peaks are slightly overestimated
by $\sim 0.3$ eV with respect to IXS data of Ref.~\onlinecite{loa_plasmons_2011},
showing a better agreement with the IXS data at 43 GPa 
of Ref.~\onlinecite{Mao20122011}.
The deviation is barely larger than the 
energy resolution (0.6 eV) of the IXS data of Ref~\onlinecite{loa_plasmons_2011}.
Furthermore, the agreement between the calculated and experimental slopes
at these two pressures indicates the adequacy of our calculations
for describing
the collective electronic properties of the system at different pressures.

\begin{figure}[t]
\centering
\includegraphics[width=0.49\textwidth]{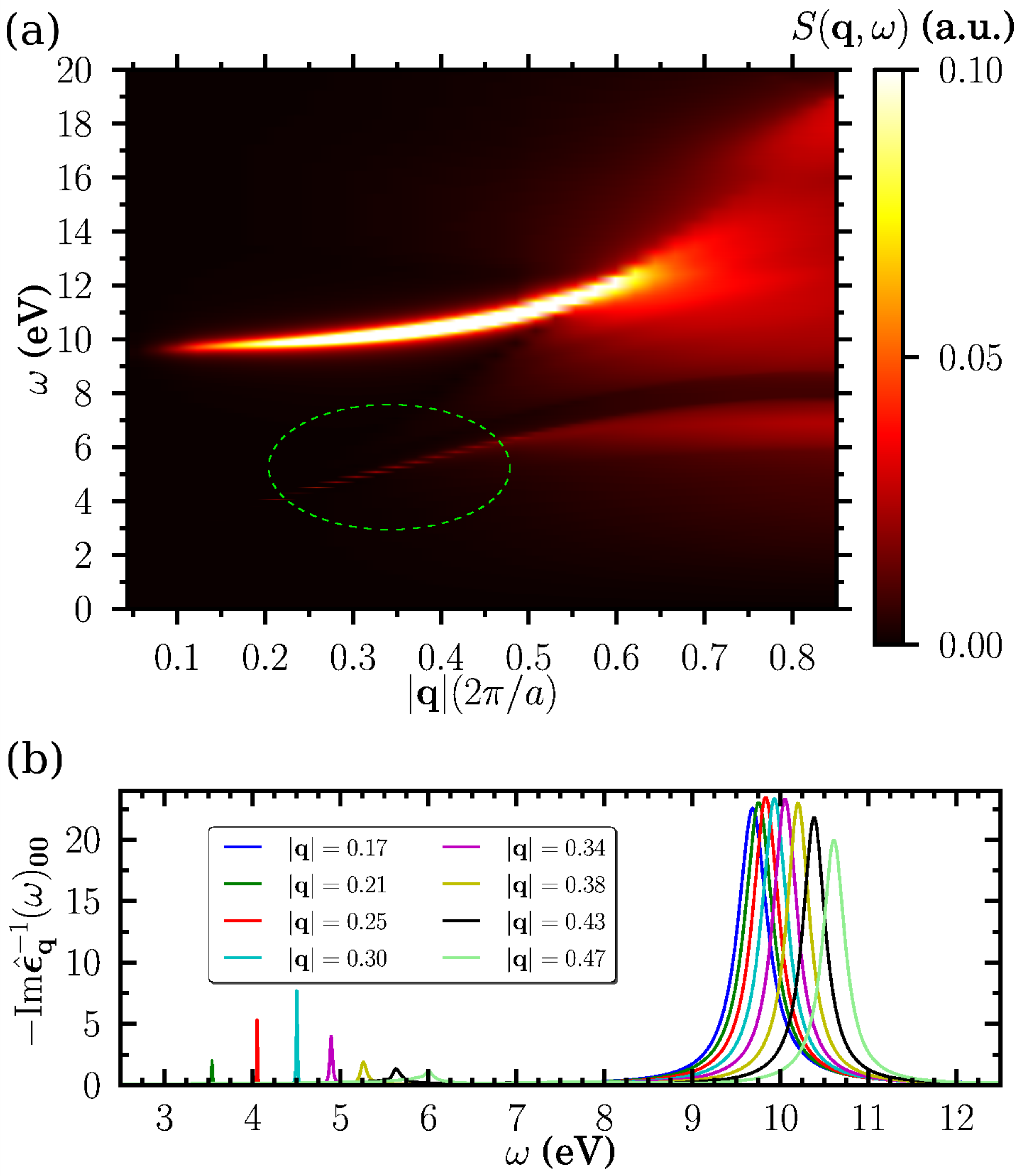}
\caption{(color online)  (a) and (b) show the dynamical structure factor 
and energy-loss function
of fcc Na at 75 GPa along the
$\Gamma$L direction, respectively. 
In (a), the dashed (green) circle encloses the area 
where the anisotropic interband plasmon emerges.
In (b), the values of the momentum 
considered are depicted in the inset 
(units of $2\pi/a$).}
\label{fig:fcc-all-plasmons-75GPa}
\end{figure}

\subsubsection{The fcc phase (65-105 GPa)}
\label{sec:fcc}

\begin{figure}[t]
\centering
\includegraphics[width=0.5\textwidth]{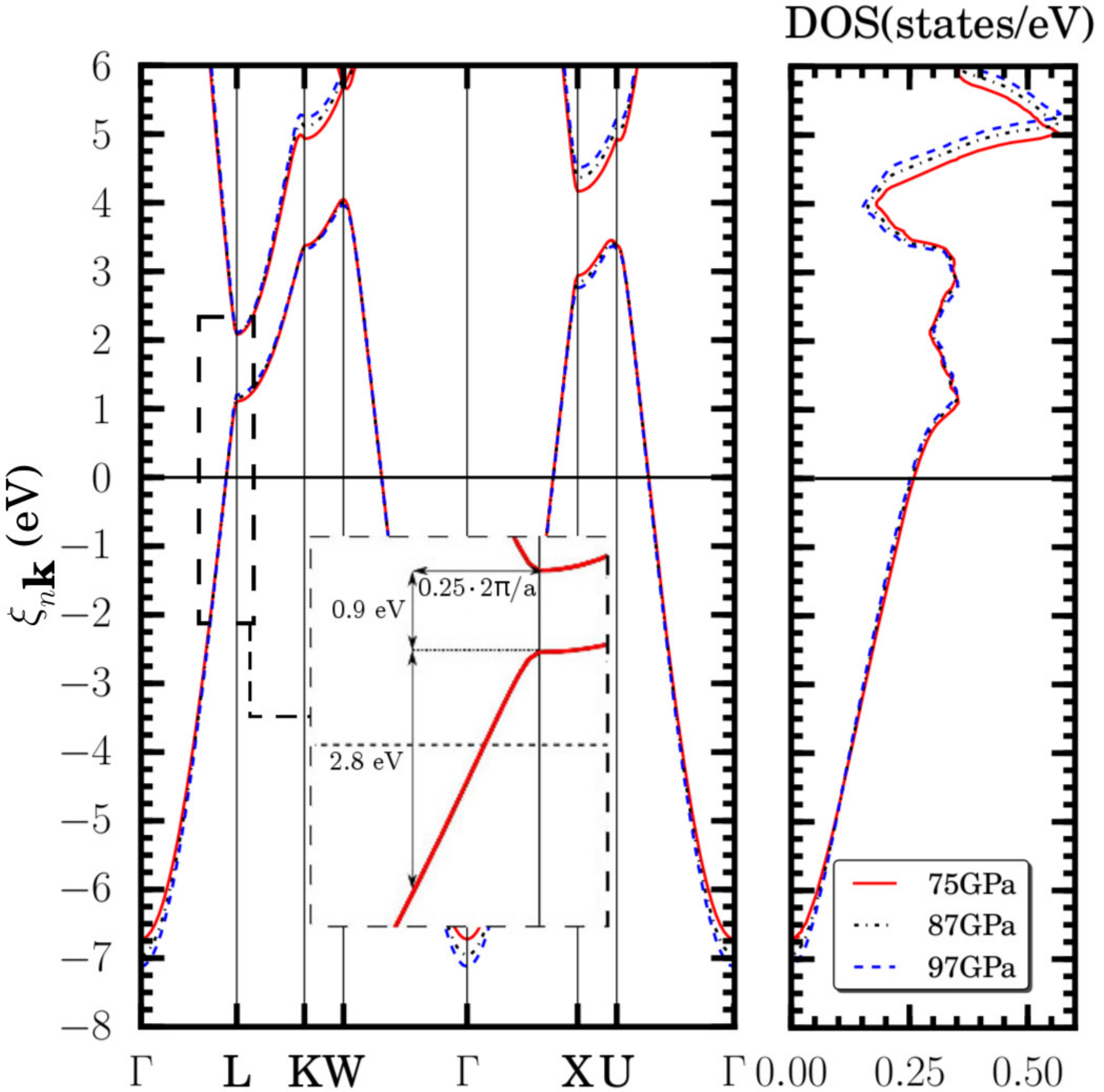}
\caption{(color online)  (Left panel) Electronic band structure of fcc Na 
at 75, 87 and 97 GPa.
The inset shows the details of the band gap around high symmetry
point L at 75 GPa. The Fermi level is indicated by the 
horizontal solid (black) line.
(Right panel) Total DOS (states/eV).
}
\label{fig:fcc-bands}
\end{figure}

At 65 GPa sodium undergoes a phase transformation from the bcc to the fcc structure. 
Unlike the rest of alkali metals, 
the Fermi surface of Na remains
spherical up to $\sim105$ GPa~\cite{alkali}.
Therefore, fcc Na can still be regarded as a simple metal.
However, our calculations characterize an anisotropic interband plasmon
along the $\Gamma$L direction, indicating
a significant departure of fcc Na
from the simple metal behavior. 

In Fig. \ref{fig:fcc-all-plasmons-75GPa}a we display
the calculated dynamical structure factor of fcc Na 
at 75 GPa along $\Gamma$L, 
showing a plasmon with parabolic dispersion
that emerges at around $9.5$ eV, in reasonable agreement
with IXS data of Ref.~\onlinecite{Mao20122011}, $\omega_{p}\simeq9.25$ eV. 
The evolution of the plasmon with respect to the momentum in the fcc phase
has been included in Fig. \ref{fig:pl-disp-bcc} for 75, 87 and 97 GPa.
Apart from showing a remarkable agreement 
with IXS data of Ref.~\onlinecite{Mao20122011}, our calculations confirm that the
plasmon dispersion is almost parabolic all over the stability pressure range
of fcc Na.

It is noteworthy that the calculated dynamical structure factor
depicted in Fig. \ref{fig:fcc-all-plasmons-75GPa}a
reveals a weaker second plasmon branch 
which does not follow at all the free-electron-like
parabolic dispersion (see inside the dashed circle).
The analysis of the energy-loss function
(Fig. \ref{fig:fcc-all-plasmons-75GPa}b) shows that
this second branch emerges and disappears at finite
values of the momentum, $|\textbf{q}|\sim 0.2\cdot2\pi/a$ and $|\textbf{q}|\sim 0.5\cdot2\pi/a$, respectively.
Furthermore, the vanishingly small plasmon linewidth
of Fig. \ref{fig:fcc-all-plasmons-75GPa}b
indicates that the associated collective excitation 
is practically undamped.

\begin{figure}[t]
\centering
\includegraphics[width=0.49\textwidth]{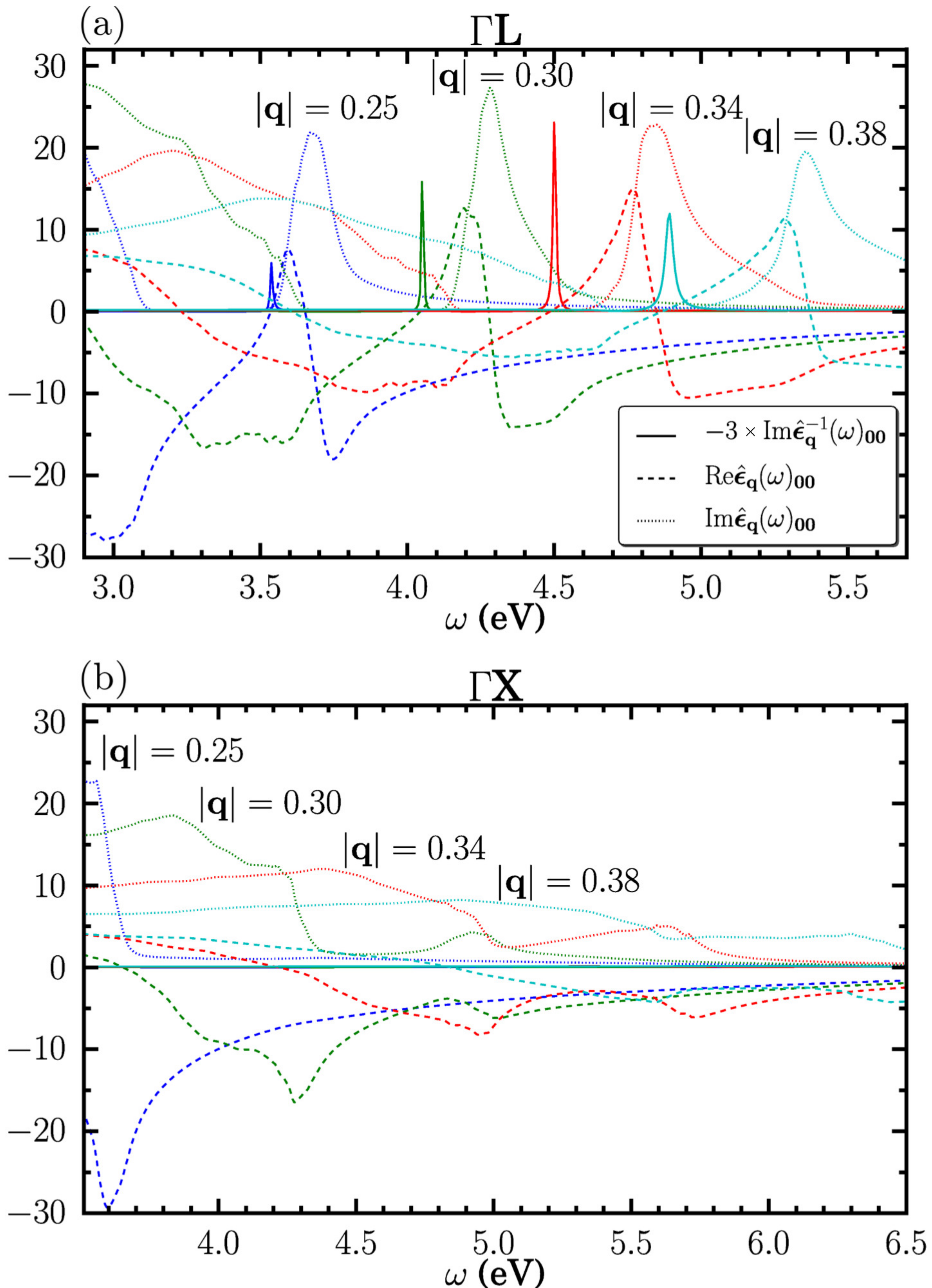}
\caption{(color online)   Real (dashed lines), 
imaginary (dotted lines) and inverse imaginary part (solid lines)
of the dielectric function of fcc Na at 75 GPa along $\Gamma$L (a)
and $\Gamma$X (b). Note the different scale for $-\Im \hat{\bm{\epsilon}}_{\textbf{q}}^{-1}(\omega)_{\textbf{00}}$. 
Shown results are for
$|\textbf{q}|=$ 0.25, 0.30, 0.34 and 0.38 (units of $2\pi/a$).
}
\label{fig:fcc-epsilon-GL-GX}
\end{figure}

In order to identify the nature of this plasmon, 
we have performed an analysis
of the energy-loss function along
different directions, 
and we have not found any similar peak of
$-\Im \hat{\bm{\epsilon}}_{\textbf{q}}^{-1}(\omega)_{\textbf{00}}$
in any other direction.
This fact reveals a strong anisotropy of the system that follows from
the electronic band structure of fcc Na depicted in Fig. \ref{fig:fcc-bands},
which also displays important anisotropic features.
Specifically, the free-electron-like band 
presents a gap at significantly different energies for different directions;
the gap opens at $\sim 1$ eV along $\Gamma$L, while for the 
rest of directions it opens around $3$ to $4$ eV. 
As shown in the next paragraph, the band structure has a direct impact on the
electron-hole excitations and is the origin of the 
anisotropic plasmon we have found.

In Fig. \ref{fig:fcc-epsilon-GL-GX} we display the real
and imaginary parts of the dielectric function
along $\Gamma$L and $\Gamma$X
for various values of the momentum.
Whereas for a given direction the calculations for different 
$\textbf{q}$'s share similar 
features,
the results along
$\Gamma$L and $\Gamma$X
exhibit important differences.
We first analyze the results along $\Gamma$L (Fig. \ref{fig:fcc-epsilon-GL-GX}a).
Focusing on $|\textbf{q}|=0.25\cdot2\pi/a$, we observe a decrease
of $\Im \hat{\bm{\epsilon}}_{\textbf{q}}(\omega)_{\textbf{00}}$ until 
it completely vanishes at $\sim$3.0 eV.
This value coincides
approximately with the 
energy at which intraband excitations along $\Gamma$L vanish
due to the opening of a band gap 
(see inset of Fig. \ref{fig:fcc-bands}).
The absence of electron-hole excitations
remains up to $\sim$3.6 eV, where 
interband transitions begin; again, this
energy coincides with the end of the 
band gap along $\Gamma$L.
The strength of interband excitations
is evidenced by the prominence of the peak at $\sim$3.7 eV
in $\Im \hat{\bm{\epsilon}}_{\textbf{q}}(\omega)_{\textbf{00}}$.
Due to Kramers-Kronig relations~\cite{pines},
the peak in the imaginary part
drives the real part to the positive side,
passing through zero at $\sim$3.6 eV and
giving rise to the interband plasmon at that energy.

The above described situation remains very similar for 
$|\textbf{q}|=$0.30, 0.34 and 0.38$\cdot2\pi/a$, 
the only relevant
difference being an overall 
shift of all the features to higher energies, 
including the plasmon peak (see Fig. \ref{fig:fcc-epsilon-GL-GX}a).
For even higher momenta (results not shown), we find 
that the intraband and interband 
excitations overlap so that 
$\Im \hat{\bm{\epsilon}}_{\textbf{q}}(\omega)_{\textbf{00}}$
does not completely vanish in the intermediate 
energy region, 
leading to a significant broadening and 
weakening of the plasmon peak.
At low momenta ($|\textbf{q}| < 0.2\cdot2\pi/a$), the interband
transitions are not sufficiently strong
for driving $\Re \hat{\bm{\epsilon}}_{\textbf{q}}(\omega)_{\textbf{00}}$ to 
the positive part
and, therefore,
we do not find any plasmon peak in the energy-loss function.

The calculated results along $\Gamma$X, illustrated in Fig. \ref{fig:fcc-epsilon-GL-GX}b,
display two major differences with respect to the ones in 
Fig. \ref{fig:fcc-epsilon-GL-GX}a.
First, the intraband excitations 
end at significantly higher energies
than in Fig. \ref{fig:fcc-epsilon-GL-GX}a due to
the absence of band gaps 
in the band structure up to $\sim$3 eV from the Fermi energy
(see Fig. \ref{fig:fcc-bands}).
As a consequence, the intraband 
and interband excitations
overlap even for $|\textbf{q}|=0.25\cdot2\pi/a$,
preventing $\Im \hat{\bm{\epsilon}}_{\textbf{q}}(\omega)_{\textbf{00}}$
from vanishing.
The second major difference resides
in the strength of the interband excitations,
which is much weaker along the $\Gamma$X direction and 
is reflected by the relative decrease of
the interband peak of $\Im \hat{\bm{\epsilon}}_{\textbf{q}}(\omega)_{\textbf{00}}$ 
as compared to that along $\Gamma$L. 
As a consequence, Kramers-Kronig relations do not drive
$\Re \hat{\bm{\epsilon}}_{\textbf{q}}(\omega)_{\textbf{00}}$
to the positive part in the 3-7 eV energy range.
Thus, unlike along $\Gamma$L, 
we do not find any interband plasmon along this or any other
highly symmetric direction.

We have performed the same analysis at 87 GPa and 97 GPa,  
verifying that the anisotropic plasmon
along $\Gamma$L persists at these pressures as well. This is 
consistent with the associated band structures
(see Fig. \ref{fig:fcc-bands}),
which share very similar features as pressure increases in the fcc
structure,
including the band gap along
$\Gamma$L at $\sim$1 eV above the Fermi level.
Therefore, our calculations suggests 
that in besides the free-electron-like plasmon
at $\sim$ 10 eV, fcc Na presents an additional
interband plasmon at $\sim$3.5-5.5 eV all over its stability
pressure range.

\begin{figure}[t]
\centering
\includegraphics[width=0.49\textwidth]{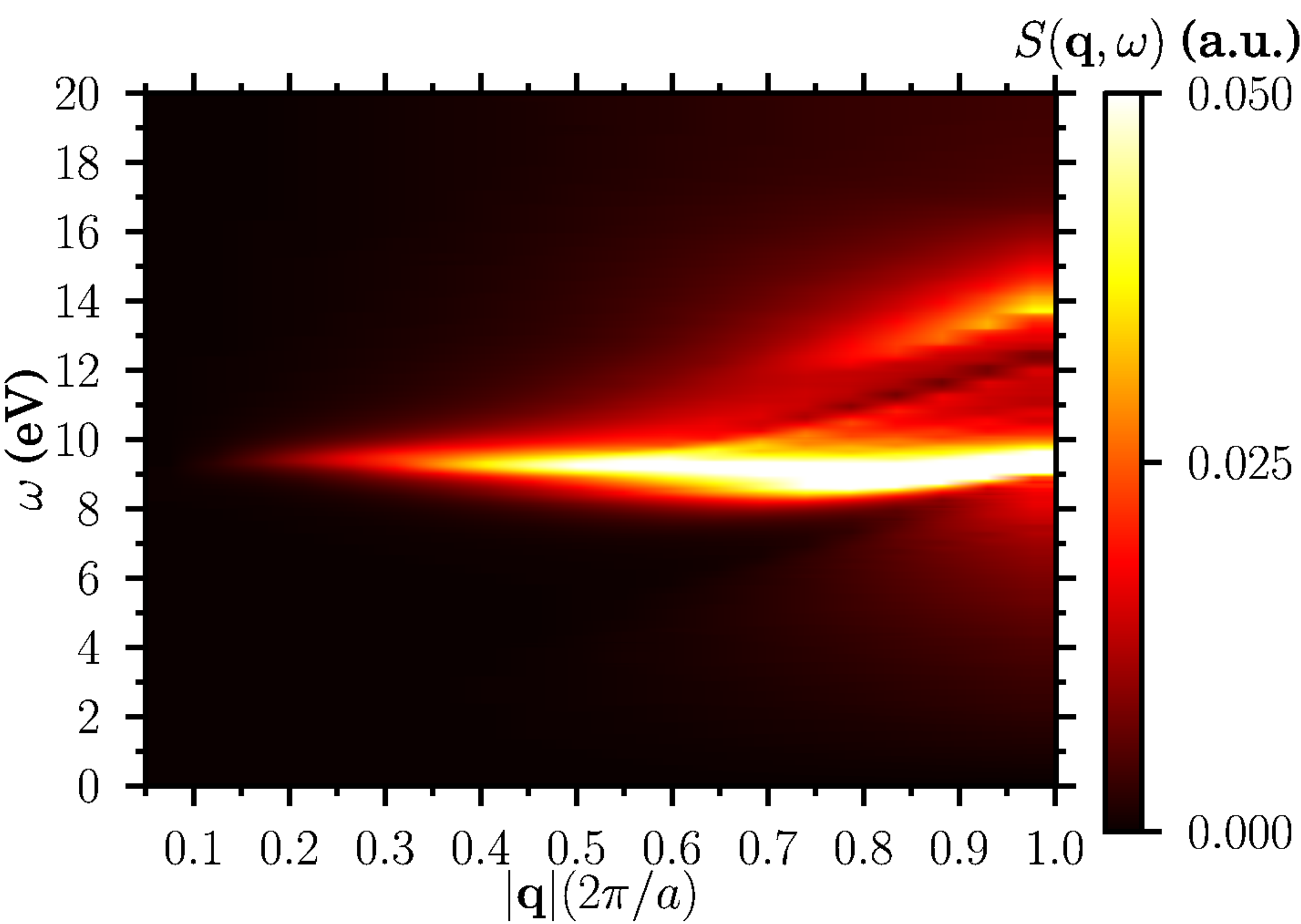}
\caption{(color online)  Dynamical structure factor 
of cI16 Na along $\Gamma$P at 105 GPa.}
\label{fig:cI16-dyn-str}
\end{figure}

\subsection{Complex phases of sodium}
\label{sec:na-complex-phases}

As characterized by several high pressure 
experiments~\cite{poster,structural-science,ma_transparent_2009},
above 105 GPA sodium adopts considerably more complex
structures than the previous bcc and fcc phases. 
Additionally, it exhibits clear fingerprints of 
pressure-induced complexity in this regime.
As an example, the reflectivity of Na has been measured to 
drastically drop at low frequencies~\cite{anomalous-pnas}, indicating a clear departure from the
expected free-electron-like behavior.  

The connection between the reflectivity and the dielectric
function of a material is given by  
\begin{equation}
\label{eq:reflectivity}
R(\omega)=\frac{(1-n(\omega))^{2}+\kappa^{2}(\omega)}
{(1+n(\omega))^{2}+\kappa^{2}(\omega)},
\end{equation}
with $n(\omega)=\Re \sqrt{\hat{\bm{\epsilon}}_{\textbf{q}}(\omega)_{\textbf{00}}}$ and 
$\kappa(\omega)=\Im\sqrt{\hat{\bm{\epsilon}}_{\textbf{q}}(\omega)_{\textbf{00}}}$.
In this section, we will analyze the evolution of this quantity 
as a function of pressure.

\subsubsection{The cI16 phase (105-118 GPa)}
\label{sec:cI16}

From 105 to 118 GPa, sodium adopts the  
cI16 structure.
Interestingly, 
we have not found any anisotropic interband plasmon 
in this pressure range.
This is exemplified in Fig. \ref{fig:cI16-dyn-str}, where the calculated 
dynamical structure factor along the $\Gamma$P direction
shows a single intraband plasmon at 
around 9.7 eV, which is 
$\sim15\%$
lower than the 
one predicted by the free-electron model.
Furthermore, our calculations 
indicate that the evolution of the plasmon dispersion
with respect to the momentum is not 
parabolic; 
in fact, Fig. \ref{fig:cI16-dyn-str} shows it is almost 
momentum-independent along $\Gamma$P.

\begin{figure}[t]
\centering
\includegraphics[width=0.49\textwidth]{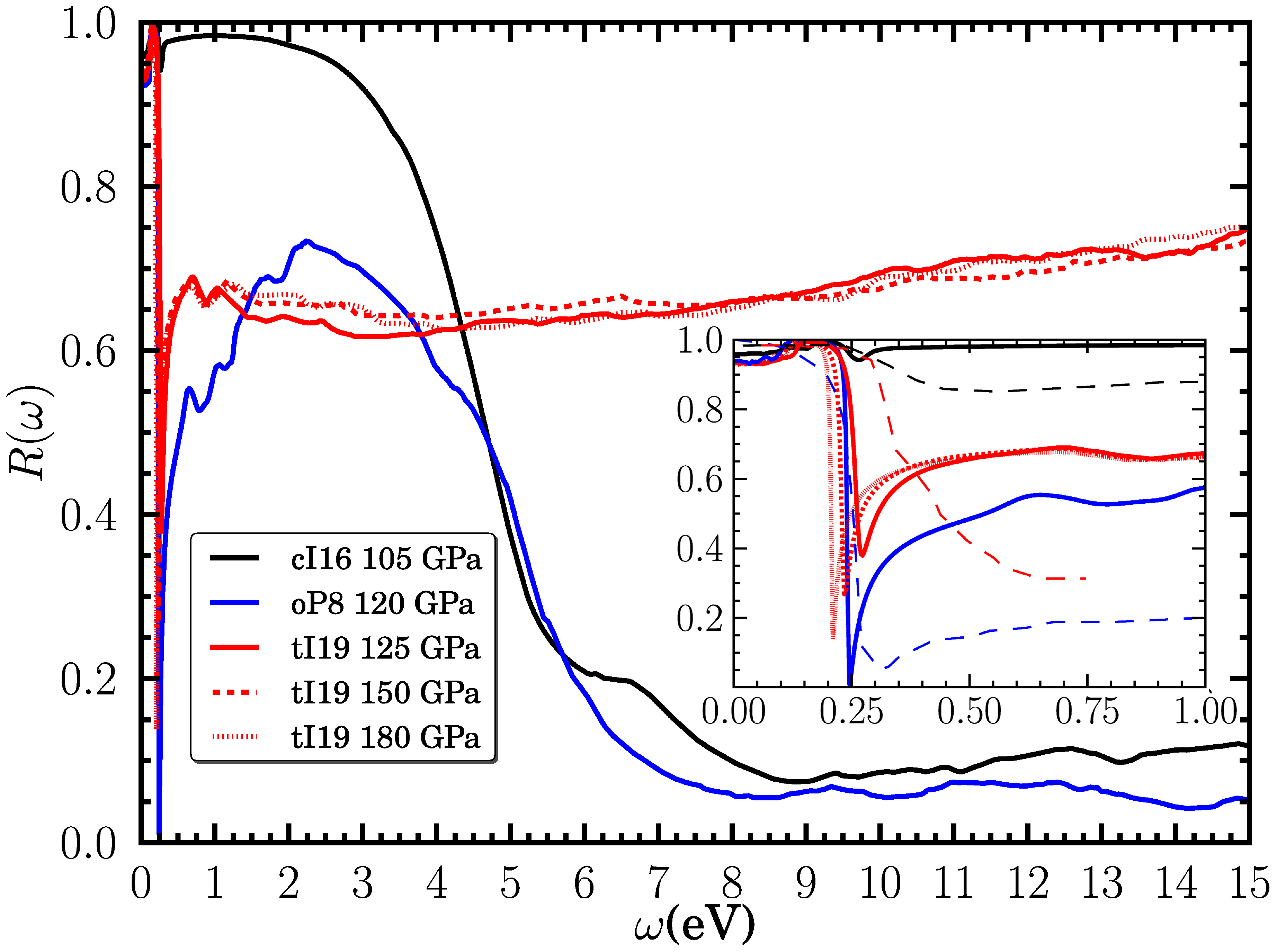}
\caption{(color online)  The calculated reflectivity spectrum of sodium in the phases 
cI16 (105 GPa), oP8 (120 GPa) and tI19 (125, 150 and 180 GPa).
The inset addresses the 0-1.0 eV range, where we have also included 
synchrotron infrared spectroscopy data (long dashed lines) from Ref. \onlinecite{anomalous-pnas}.}
\label{fig:reflectivity}
\end{figure}

In Fig. \ref{fig:reflectivity} we show the calculated
reflectivity (Eq. \ref{eq:reflectivity}) for cI16 Na.
Our results indicate 
an almost complete light reflection
from 0 to 3 eV.
This property is in reasonable
agreement with recent experiments~\cite{anomalous-pnas} 
measuring a constant reflectivity $R(\omega)\simeq 0.85$ over the same frequency range.
At 3 eV, the reflectivity
starts a smooth decrease that 
ends at around 10 eV, where $R(\omega)$ is practically
suppressed as a consequence of the intraband plasmon.

\subsubsection{The oP8 phase (118-125 GPa)}
\label{sec:oP8}

Beyond 118 GPa, after a phase transformation favoring
the oP8 structure~\cite{structural-science}, sodium exhibits an anomalous 
behavior associated to its optical response.
This fact is clearly exemplified by Fig. \ref{fig:reflectivity}, where the
calculated reflectivity (Eq. \ref{eq:reflectivity}) in the oP8 phase shows a sudden
dip at around 0.25 eV,
vanishing almost completely.
This behavior is in remarkable quantitative agreement with 
recent optical measurements
showing a drop of the reflectivity to 0.05 at practically
the same energy~\cite{anomalous-pnas} (see inset of Fig. \ref{fig:reflectivity}). 
After the dip, 
the reflectivity of oP8 Na partially recovers, 
but around 2 eV it starts a smooth 
decrease until $\sim 7$ eV, where it 
becomes almost zero.

\begin{figure}[t]
\centering
\includegraphics[width=0.49\textwidth]{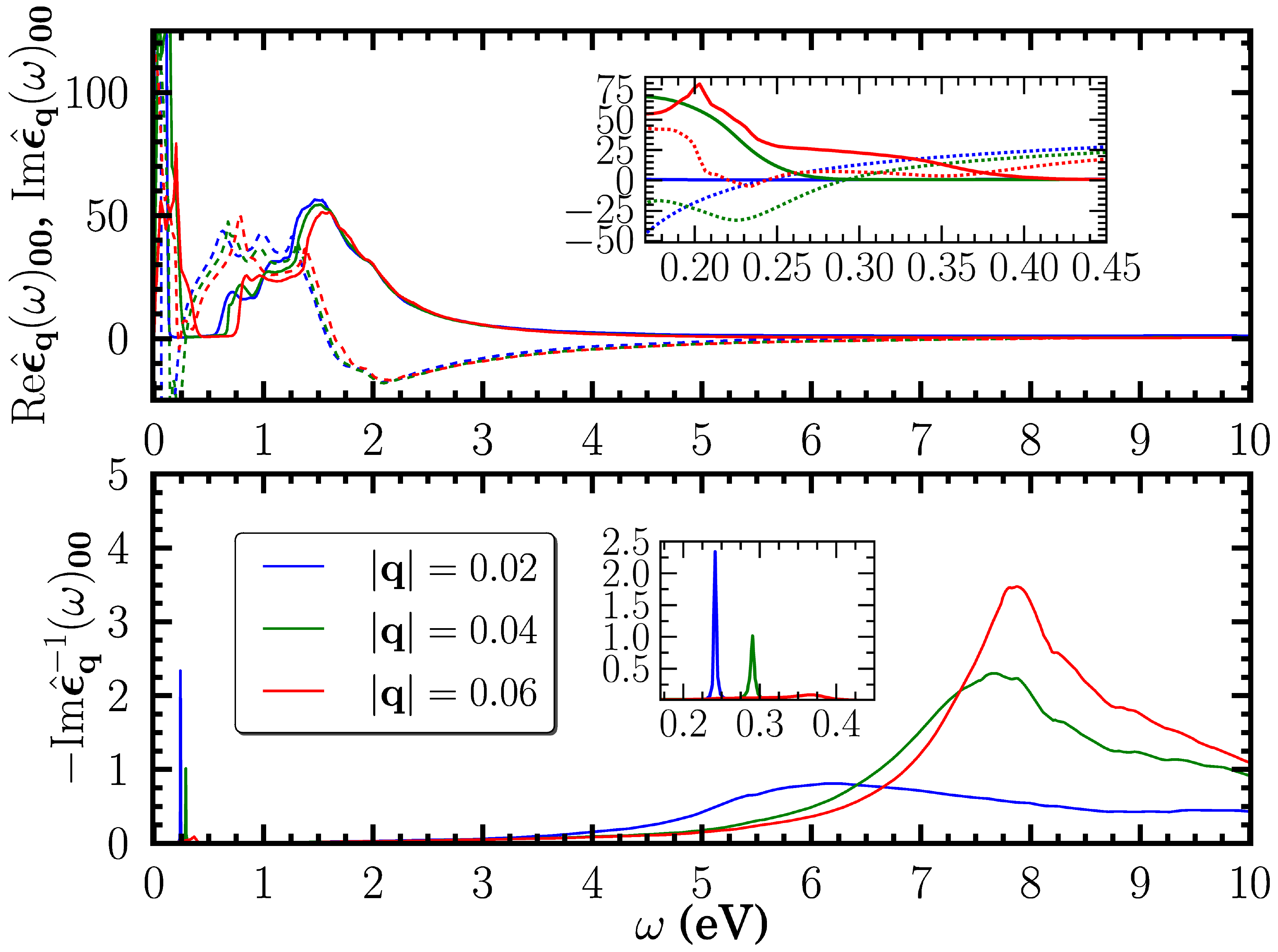}
\caption{(color online)  Dielectric function of oP8 Na at 120 GPa.
Top panel: 
$\Im \hat{\bm{\epsilon}}_{\textbf{q}}(\omega)_{\textbf{00}}$ and 
$\Re \hat{\bm{\epsilon}}_{\textbf{q}}(\omega)_{\textbf{00}}$ 
are depicted by solid and dashed lines, respectively.
Bottom panel: $-\Im \hat{\bm{\epsilon}}^{-1}_{\textbf{q}}(\omega)_{\textbf{00}}$. 
In both panels, 
results are shown for $|\textbf{q}|=$0.02, 0.04 and $0.06\cdot2\pi/a$.
Both insets illustrate the 0-0.5 eV range.}
\label{fig:op8-epsilon}
\end{figure}

In Fig. \ref{fig:op8-epsilon} we analyze 
the dielectric function of oP8 Na. 
This figure reveals that the anomalous behavior of 
$R(\omega)$ in the optical range originates
from a low-energy plasmon
emerging at around 0.25 eV.
This plasmon shares common features with 
the one theoretically predicted in calcium under pressure~\cite{calcium-ion},
which also induces a dip in the calculated reflectivity. 
It is essentially undamped since both $\Im \hat{\bm{\epsilon}}_{\textbf{q}}(\omega)_{\textbf{00}}$ 
and $\Re \hat{\bm{\epsilon}}_{\textbf{q}}(\omega)_{\textbf{00}}$ become
almost zero at $\omega\simeq 0.25$ eV, 
making the plasmon linewidth vanishingly small at this
energy. 
For increasing values of the momentum, 
the linewidth of the low-energy plasmon
starts broadening until $|\textbf{q}|=0.06\cdot2\pi/a$,
where the peak in the energy-loss function is 
practically suppressed. 
At higher energies, our calculations 
evidence the existence of an
intraband plasmon at around 6$-$8 eV that
coincides with the final loss of reflectivity
depicted in Fig. \ref{fig:reflectivity}.

As shown in the top panel of Fig. \ref{fig:op8-epsilon},
for $|\textbf{q}|< 0.04\cdot2\pi/a$
our calculations evidence
a gap between the intraband and interband 
excitations in the $\sim$0-0.6 eV energy range, 
where $\Im \hat{\bm{\chi}}^\text{KS}_{\textbf{q}}(\omega)_{\textbf{00}}$ 
completely vanishes. 
Driven by the Kramers-Kronig relations,
the gap and the subsequent interband excitations 
contributing to the imaginary part 
($\omega > 0.6$ eV)
force the real part of the dielectric function to become positive  
at $\omega\simeq 0.25$ eV,
giving rise to the low-energy interband plasmon at that energy.

\begin{figure}[t]
\centering
\includegraphics[width=0.45\textwidth]{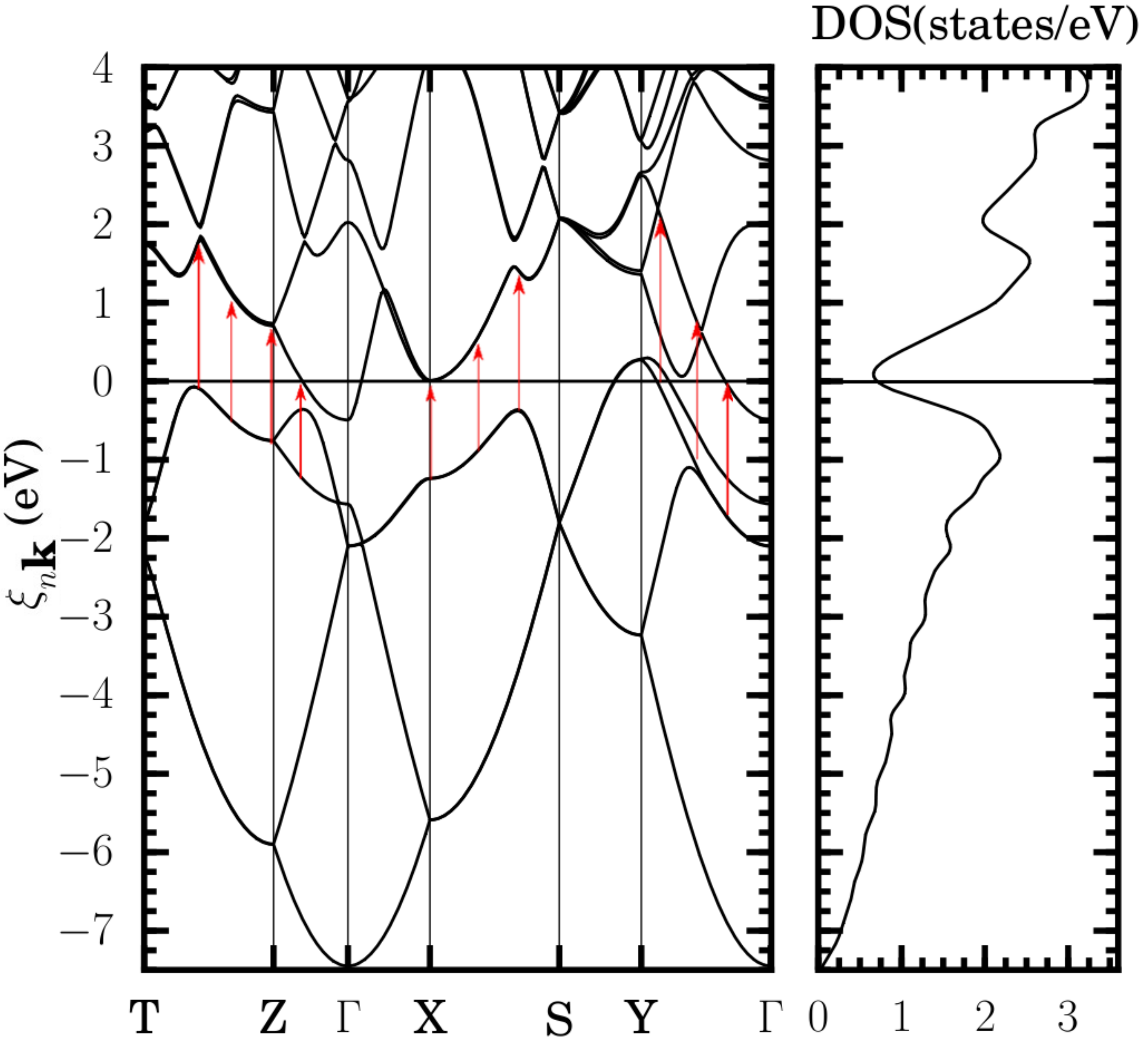}
\caption{(color online)  (Left panel) Electronic band structure of oP8 Na at 120 GPa.
Red arrows depict the interband transitions responsible for
the emergence of the low-energy plasmon. The Fermi level is indicated by the 
horizontal solid (black) line.
(Right panel) Total DOS (states/eV).
}
\label{fig:oP8-bands-dos}
\end{figure}

The relevant interband excitations 
contributing to $\Im \hat{\bm{\epsilon}}_{\textbf{q}}(\omega)_{\textbf{00}}$
in the \textbf{q}$\rightarrow$0 limit are 
characterized in the band structure of oP8 Na depicted in
Fig. \ref{fig:oP8-bands-dos}, 
alongside with the calculated DOS, 
which reproduces the weakening of the metallic 
character at the Fermi level reported in other works~\cite{anomalous-pnas,na-calc}.
As indicated by the red arrows in Fig. \ref{fig:oP8-bands-dos}, 
there exist quasi-parallel occupied-unoccupied
bands separated by 1-2 eV along various directions
in reciprocal space: ZT, $\Gamma$Z, XS and $\Gamma$Y,
among others.
The energy difference between these bands
coincides with the
interband excitations contributing to $\Im \hat{\bm{\epsilon}}_{\textbf{q}}(\omega)_{\textbf{00}}$
for $\omega\sim 1-2$ eV (see top panel of Fig. \ref{fig:op8-epsilon}), and are 
therefore directly responsible for the
emergence of the low-energy plasmon at that energy.

\subsubsection{The tI19 phase (125-180 GPa)}
\label{sec:tI19}
\begin{figure}[t]
\centering
\includegraphics[width=0.49\textwidth]{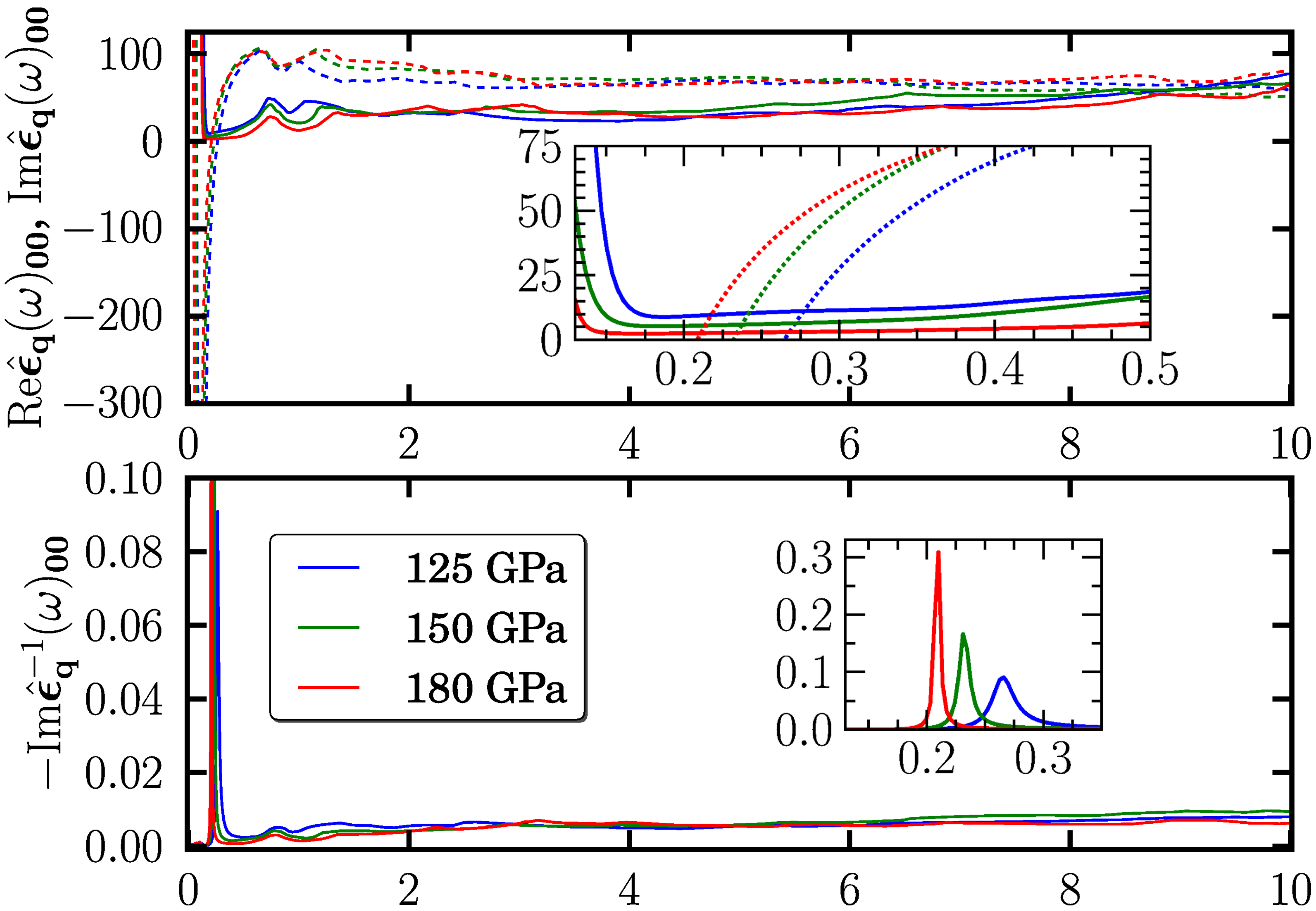}
\caption{(color online)  Dielectric function of tI19 Na at 125, 150 and 180 GPa
for $|\textbf{q}|=0.03\cdot2\pi/a$ along $\Gamma$X.
Top panel: 
$\Im \hat{\bm{\epsilon}}_{\textbf{q}}(\omega)_{\textbf{00}}$ and 
$\Re \hat{\bm{\epsilon}}_{\textbf{q}}(\omega)_{\textbf{00}}$ 
are depicted by solid and dashed lines, respectively.
Bottom panel: $-\Im \hat{\bm{\epsilon}}^{-1}_{\textbf{q}}(\omega)_{\textbf{00}}$.}
\label{fig:tI19-epsilon}
\end{figure}

At 125 GPa sodium adopts the tI19 structure~\cite{structural-science}.
As in the oP8 phase, we have also characterized a very low-energy
plasmon, shown in Fig. \ref{fig:tI19-epsilon}, that induces a sudden dip on the 
optical reflectivity at around 0.25 eV (see Fig. \ref{fig:reflectivity}),
in qualitative agreement with optical measurements~\cite{anomalous-pnas}.
We find two major differences between the reflectivity spectrum 
of the oP8 and tI19 phases. First, the minimum of
$R(\omega)$ ranges from $\sim 0.4$ to $\sim 0.2$
throughout the stability pressure range of the tI19 phase, 
whereas in the oP8 phase $R(\omega)\simeq 0$ at the dip. 
Second, unlike in the case of oP8 Na, the reflectivity of tI19 Na is almost
totally recovered in the infrared regime, i.e. $R(\omega)\geq 0.6$ for 
$\omega>0.75$ eV. 
These experimentally supported features~\cite{anomalous-pnas} 
indicate that tI19 Na shows better metallic properties than oP8 Na, 
excluding a possible metal-insulator transition between the two phases.

\begin{figure}[t]
\centering
\includegraphics[width=0.45\textwidth]{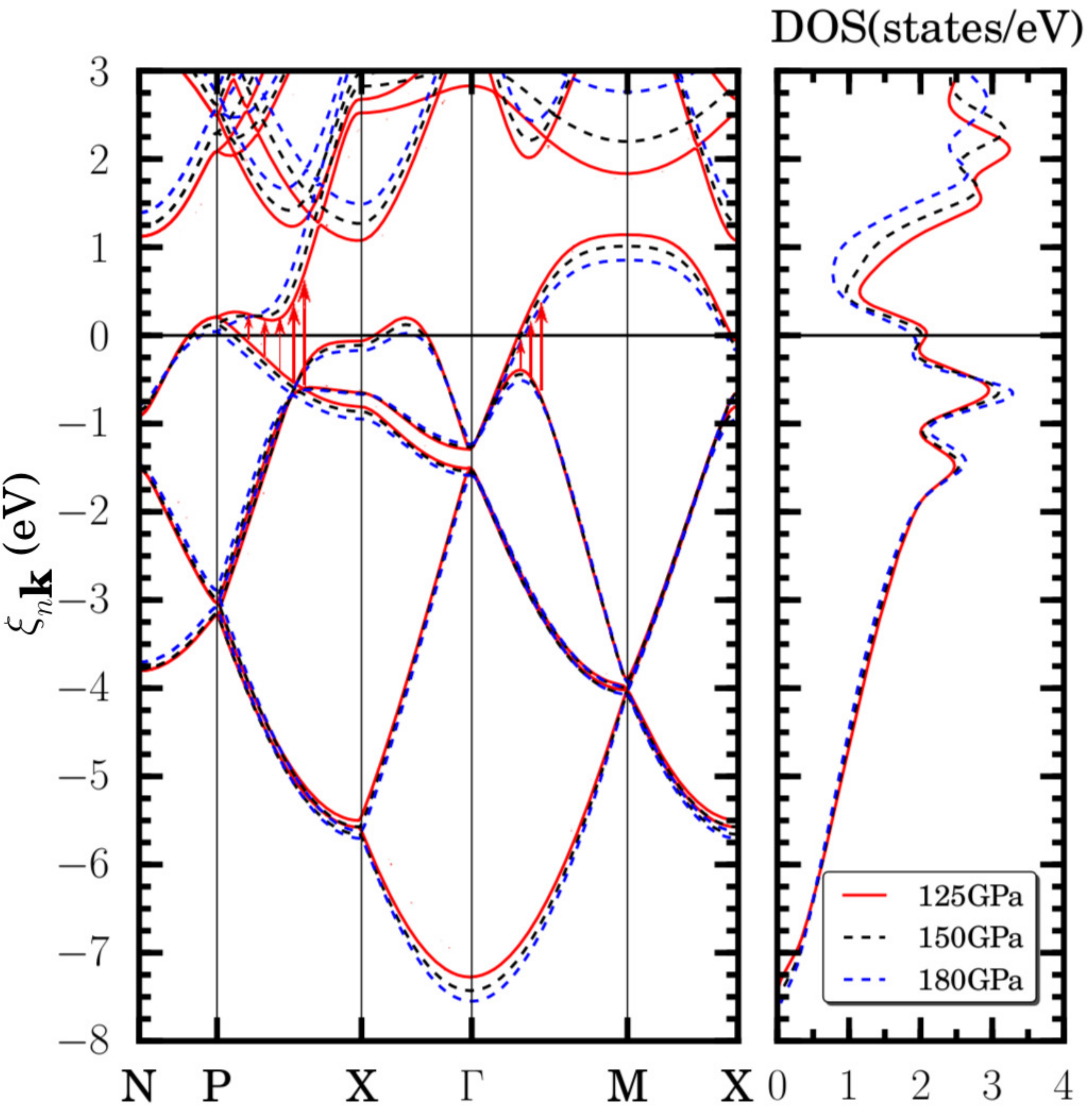}
\caption{(color online) (Left panel) Electronic band structure of tI19 Na at 125, 150 and 180 GPa.
Red arrows depict the interband transitions responsible for
the emergence of the low-energy plasmon. The Fermi level is indicated by the 
horizontal solid (black) line.
(Right panel) Total DOS (states/eV).}
\label{fig:tI19-band-structure}
\end{figure}

We have performed an analysis of the dielectric response function throughout the 
stability pressure range of tI19 Na from 125 to 180 GPa, as shown in Fig. \ref{fig:tI19-epsilon}.
We find that the low-energy plasmon  persists over all the studied domain. 
Moreover, our calculations indicate
that the plasmon becomes undamped as pressure is increased, i.e.
its linewidth decreases with increasing pressure (see Fig. \ref{fig:tI19-epsilon}b). 
As shown in the inset of Fig. \ref{fig:tI19-epsilon}a,  
$\Im \hat{\bm{\epsilon}}_{\textbf{q}}(\omega)_{\textbf{00}}$
does not completely vanish in the $\sim0.2-0.5$ eV range, where
$\Re \hat{\bm{\epsilon}}_{\textbf{q}}(\omega)_{\textbf{00}}$
becomes vanishingly small. As a consequence, the resulting plasmon acquires
a finite linewidth. Since 
$\Im \hat{\bm{\epsilon}}_{\textbf{q}}(\omega)_{\textbf{00}}$
decreases (approaches zero) with increasing pressure, so does
the plasmon linewidth. 

\begin{figure*}[t]
\centering
\includegraphics[width=0.7\textwidth]{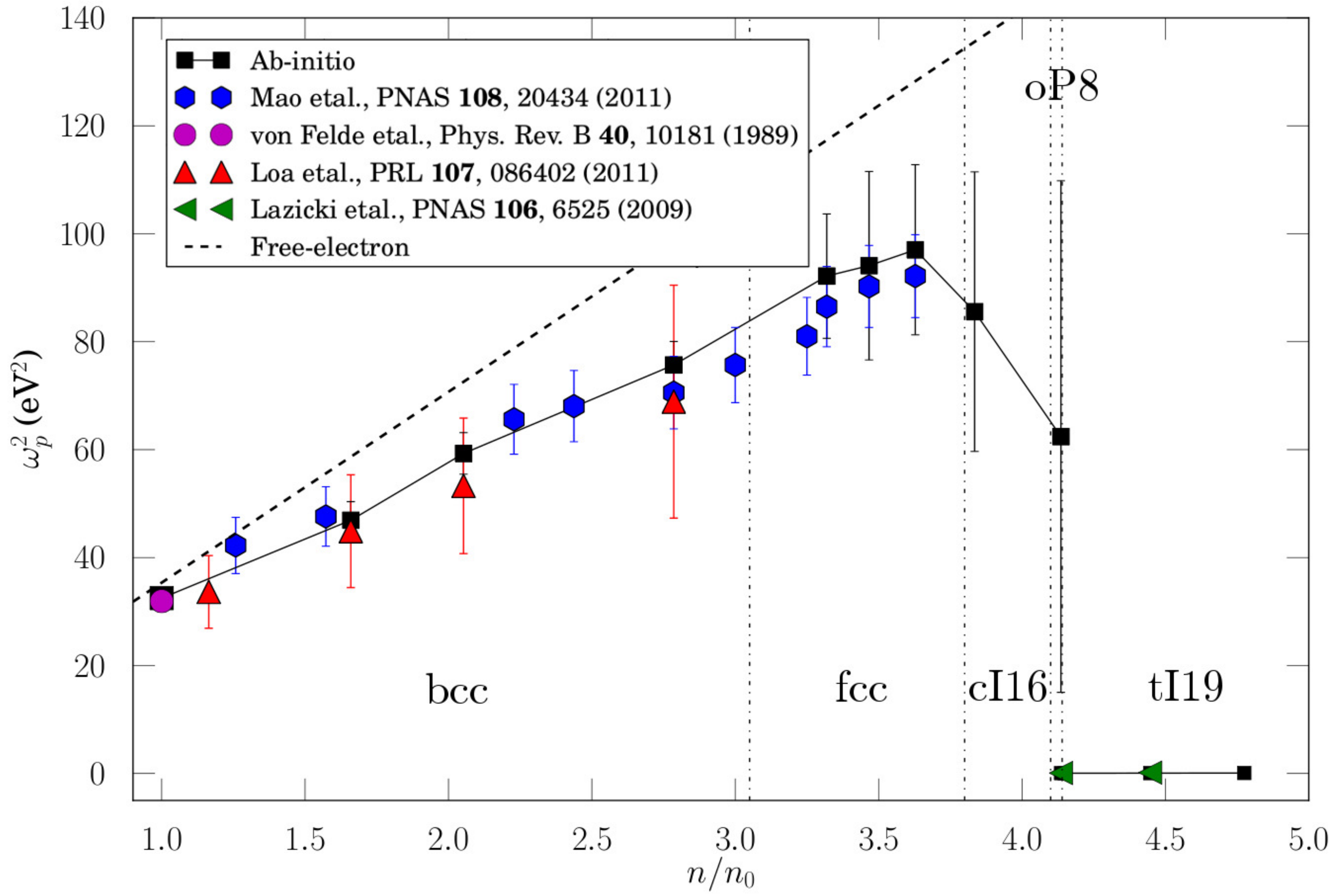}
\caption{(color online) Plasmon dispersion of Na at \textbf{q}$\rightarrow0$ from 0 to 180 GPa 
with respect to the density. 
Squares (black) represent our ab-initio calculations including the plasmon linewidth, 
and solid lines are simple guides to the eye. Experimental data are taken from Refs.
~\onlinecite{bcc-0gpa,loa_plasmons_2011,Mao20122011,anomalous-pnas}.
Vertical dot-dashed lines separate different phases of Na.
The dashed line denotes the free-electron dispersion, 
$\omega_{p}^{2}=4\pi n/m^{2}$.
}
\label{fig:plasmon-density}
\end{figure*}

In Fig. \ref{fig:tI19-band-structure} 
we show the calculated electronic band structure and DOS
of tI19 Na at 125, 150 and 180 GPa. 
The electronic excitations contributing to 
$\Im \hat{\bm{\epsilon}}_{\textbf{q}}(\omega)_{\textbf{00}}$ in the
$0.5-1.0$ eV range (see Fig. \ref{fig:tI19-epsilon}a) are characterized by red arrows,
evidencing, as in the oP8 phase, the interband nature of the low-energy plasmon.  
Another important detail revealed by Fig. \ref{fig:tI19-band-structure} 
is that tI19 Na develops
two hole-pockets, one
around high symmetry point P and the other one halfway between $\Gamma$ and X. 
Low-energy intraband excitations 
to these hole-pockets are the reason why 
$\Im \hat{\bm{\epsilon}}_{\textbf{q}}(\omega)_{\textbf{00}}$
does not completely vanish in the $\sim0.2-0.4$ eV energy range.
Furthermore, Fig. \ref{fig:tI19-band-structure} 
indicates that the area of the hole-pockets
diminishes with increasing pressure, 
yielding weaker low-energy excitations at high pressures.
We conclude that the damping of the low-energy 
plasmon revealed by Fig. \ref{fig:tI19-epsilon}b is  
directly associated
to the evolution of the hole-pockets under pressure.

\section{CONCLUSIONS}
\label{sec:conclusions}

To conclude our analysis, 
in Fig. \ref{fig:plasmon-density} we illustrate the plasmon evolution of sodium in the optical limit 
(\textbf{q}$\rightarrow0$)
as a function of the density, covering all the phases from 0 to 180 GPa.
This figure makes it clear that sodium increasingly departs from the free-electron-like model 
as its density is raised. 
In particular, the figure evidences 
a complete breakdown of the free-electron-like 
picture at the high pressure phases cI16, oP8 and tI19.
Classically, the plasmon energy should increase with the density, but
our calculations show a marked decrease 
in the plasmon energy in the cI16 and oP8 phases,
accompanied by a huge increase of the plasmon linewidth.  
Finally,  our calculations predict the absence of intraband plasmons in the tI19 phase,
where only the low-energy interband plasmon is present. 
Therefore, sodium represents a clear example of 
how pressure can induce great complexity even in the simplest elements.

In this paper we have 
presented theoretical \textit{ab-initio}
calculations regarding the dielectric
response of bulk sodium in its 5 known
metallic phases from 0 to 180 GPa
at room temperature.
We have employed a formalism based
on a Wannier interpolation scheme 
that provides a very accurate sampling 
of reciprocal space and allows the resolution
of sharp features associated to the dielectric function. 
In this way, we predict the existence
of a low-energy plasmon
in the high pressure phases oP8 and tI19 that
explains the anomalous behavior
in the recently measured optical reflectivity,~\cite{anomalous-pnas}
also reproduced by our calculations.
The combined analysis of the KS
response function
and the electronic band structure reveals the 
interband nature of this low-energy plasmon,
associated to electron-hole transitions
between a network of quasi-parallel occupied-unoccupied bands.
Additionally,
our calculations characterize 
an anisotropic interband plasmon all along the
stability pressure range of the 
fcc configuration (65 to 105 GPa),
revealing an unexpected departure of fcc Na
from the free-electron model.
This remarkable plasmon is found exclusively
along the $\Gamma$L direction due to an anisotropic 
non-free-electron-like band structure
effect.

\section*{ACKNOWLEDGMENTS}

We are very grateful to Ion Errea for many fruitful discussions.
The authors acknowledge financial 
support from UPV/EHU (Grant No. IT-366-07) and the Spanish Ministry of 
Science and Innovation (Grant No. FIS2010-19609-C02-00).  Computer facilities 
were provided by the Donostia International Physics Center (DIPC).


\bibliographystyle{apsrev}


\end{document}